\documentclass[conference]{IEEEtran}
\IEEEoverridecommandlockouts
\usepackage{cite}
\usepackage{amsmath,amssymb,amsfonts}
\usepackage{algorithmic}
\usepackage{graphicx}
\usepackage{textcomp}
\usepackage{xcolor}
\usepackage{hyperref}
\usepackage{booktabs}
\def\BibTeX{{\rm B\kern-.05em{\sc i\kern-.025em b}\kern-.08em
    T\kern-.1667em\lower.7ex\hbox{E}\kern-.125emX}}

\begin{document}

\title{The T12 System for AudioMOS Challenge 2025: Audio Aesthetics Score Prediction System Using KAN- and VERSA-based Models\\
}

\author{\IEEEauthorblockN{Katsuhiko Yamamoto$^*$\thanks{$^*$These authors contributed equally to this work.}}
\IEEEauthorblockA{
\textit{CyberAgent}\\
Tokyo, Japan \\
{\nolinkurl{yamamoto_katsuhiko@cyberagent.co.jp}}}
\and
\IEEEauthorblockN{Koichi Miyazaki$^*$}
\IEEEauthorblockA{
\textit{CyberAgent}\\
Tokyo, Japan \\
{\nolinkurl{miyazaki_koichi_xa@cyberagent.co.jp}}}
\and
\IEEEauthorblockN{Shogo Seki$^*$}
\IEEEauthorblockA{
\textit{CyberAgent}\\
Tokyo, Japan \\
{\nolinkurl{seki_shogo@cyberagent.co.jp}}}
}

\maketitle

\begin{abstract}
We propose an audio aesthetics score (AES) prediction system by CyberAgent (AESCA) for AudioMOS Challenge 2025 (AMC25) Track 2. 
The AESCA comprises a Kolmogorov--Arnold Network (KAN)-based audiobox aesthetics and a predictor from the metric scores using the VERSA toolkit. 
In the KAN-based predictor, we replaced each multi-layer perceptron layer in the baseline model with a group-rational KAN and trained the model with labeled and pseudo-labeled audio samples. 
The VERSA-based predictor was designed as a regression model using extreme gradient boosting, incorporating outputs from existing metrics. 
Both the KAN- and VERSA-based models predicted the AES, including the four evaluation axes. 
The final AES values were calculated using an ensemble model that combined four KAN-based models and a VERSA-based model. 
Our proposed T12 system yielded the best correlations among the submitted systems, in three axes at the utterance level, two axes at the system level, and the overall average. 
We also released the inference model of the proposed KAN-based predictor (KAN \#1-\#4) \footnote{\url{https://github.com/CyberAgentAILab/aesca}}. 
\end{abstract}


\section{Introduction}
The field of generative artificial intelligence (AI) has witnessed significant advancements, particularly in the creation of synthetic audio encompassing speech, music, and other sounds. 
Because the primary users of these audio outputs are humans, evaluating the quality and effectiveness of these systems poses challenges similar to those found in the VoiceMOS challenge (VMC) for speech synthesis \cite{huang2022vmc22,Cooper2023vmc23,Huang2024vmc24}. 
To engage with the latest developments in this field, the scope of the challenge was expanded and is now recognized as the AudioMOS Challenge 2025 (AMC25)\footnote{\url{https://sites.google.com/view/voicemos-challenge/audiomos-challenge-2025}}.

This challenge includes Track 2 that focused on assessing the outputs from text-to-speech (TTS), text-to-audio (TTA), and text-to-music (TTM) systems using the aesthetic scores (AES), a novel evaluation framework \cite{Tjandra2024audiobox}. 
This framework aims to clarify evaluation by considering four distinct dimensions. 
Production Quality (PQ) examines technical attributes such as clarity and sound fidelity. 
Production Complexity (PC) assesses the intricacy of audio composition. Content Enjoyment (CE) evaluates subjective enjoyment and artistic value, whereas Content Usefulness (CU) considers the audio's potential for creative projects. 

The task of AMC25 Track 2 was to accurately predict the AES of the TTS, TTA, and TTM samples. 
The AES predicted by the submitted systems was compared with the ground truth and evaluated using the mean-squared error (MSE), linear correlation coefficient, Spearman's rank correlation coefficient (SRCC), and Kendall's rank correlation coefficient at both the utterance and system levels.

\section{Submitted \texttt{T12} System (AESCA)}

\begin{figure}
    \centering
    \includegraphics[width=0.65\linewidth]{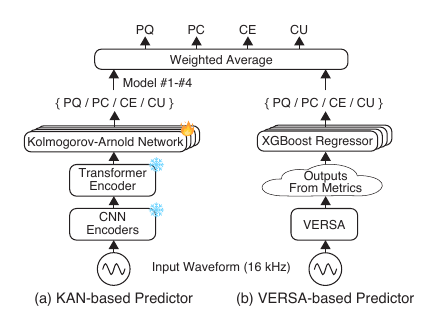}
    \vspace{-15pt}
    \caption{Overview of the proposed \texttt{T12} system for AMC25 Track 2. The system is a weighted ensemble model consisting of a KAN-based Audiobox-Aesthetics predictor (a) and a VERSA-based regression predictor (b).}
    \label{fig:proposed}
\end{figure}

Figure \ref{fig:proposed} shows the proposed \texttt{T12} system named ``AESCA (AES prediction system by CyberAgent).'' 
AESCA comprises a baseline system \cite{Tjandra2024audiobox} extended by a Kolmogorov--Arnold network (KAN) \cite{liu2025kan} and a VERSA-based predictor \cite{shi2025versa}. 
In this section, we introduce the baseline model and provide details regarding each predictor and the ensemble model. 

\subsection{Baseline Model}

Audiobox aesthetics \cite{Tjandra2024audiobox} is a reference-free AES predictor designed to assess the AES of arbitrary audio clips, including speech and music at the utterance level. 
The model was built on a pretrained WavLM encoder comprising 12 transformer layers with a hidden size of 768 \cite{Chen2022wavlm}. 
The encoder outputs branches into four independent paths, each of which applies learnable scalar weights to the outputs of each layer, aggregating information across the layers and time steps. 
These weighted audio representations were then fed into their respective multi-layer perceptrons (MLPs). 
They were responsible for regressing one of the following AES dimensions: PQ, PC, CE, or CU. 
The model was pretrained on a large dataset of approximately 500 hours, comprising 97,000 audio samples evenly distributed across speech, sound, and music. 

\subsection{KAN-based Audiobox Aesthetics}

Building upon the foundation of the baseline model, which utilizes pretrained MLPs to regress the AES dimensions, we introduce enhancements through KAN. 

KAN augments neural networks by replacing fixed activation functions with learnable rational functions for each input--output pair, enabling highly expressive nonlinear modeling \cite{liu2025kan}. 
For instance, the Kolmogorov--Arnold transformer (KAT) \cite{yang2025kat} replaces MLP layers with group-rational KAN (GR-KAN) layers, and KAT improves the image recognition performances compared with the vision transformer \cite{dosovitskiy2021vit}. 
Moreover, it is compatible with pretrained MLP weights, allowing seamless integration into existing transformer architectures without retraining, and offering improved flexibility and modeling capacity. 

In the proposed KAN-based model, we replaced each MLP layer in the baseline model with a GR-KAN. 
This is expected to enable the model to effectively utilize the parameters of the baseline model \cite{Tjandra2024audiobox}, which was pretrained on a large amount of AES data while allowing for a more flexible representation.

\subsection{VERSA-based predictor}
AES correlates significantly with various evaluation metrics, indicating the overlapping yet complementary aspects of audio quality \cite{Tjandra2024audiobox}. 
Thus, we considered a regression model that combines these metrics to support the KAN-based predictor. 

The VERSA toolkit \cite{shi2025versa} provides a range of reference-free evaluation models. 
For instance, VERSA includes DNSMOS \cite{Reddy2021dnsmos}, NISQA \cite{Mittag2021nisqa}, UTMOSv2 \cite{Baba2024utmosv2}, Torch-Squim, which predicts subjective quality scores (PESQ, STOI, and SI-SDR)\cite{Kumar2023torchaudio}, and the prompting audio-language model (PAM) for generative audio quality \cite{Deshmukh2023pam}. 
VERSA also supports the audiobox aesthetics \cite{Tjandra2024audiobox}, which is the baseline model for this challenge. 

Our VERSA-based predictor treats these evaluation metrics as input features and utilizes an extreme gradient boosting (XGBoost) \cite{Chen2016xgboost} regression model to predict the AES. 

\subsection{Ensemble model}

For the final submission, we considered an ensemble model using the stacking method of KAN- and VERSA-based predictors because an ensemble model can improve the generalization performance of each predictor. 
Four KAN-based models were trained using different random seeds \#1--\#4. 
Then, we optimized the coefficients for the four KAN-based models and one VERSA-based model using the grid search method.

\section{Experiments}

\subsection{Dataset}

We used the AMC25 dataset for Track2\footnote{\url{https://github.com/facebookresearch/audiobox-aesthetics/tree/main/audiomos2025_track2}}, including LibriTTS \cite{zen19_interspeech}, Common Voice 13.0 \cite{ardila-etal-2020-common}, EARS \cite{richter2024ears}, MUSDB18-HQ \cite{musdb18-hq}, AudioSet \cite{Gemmeke2017audioset}, and MusicCaps \cite{agostinelli2023musiccaps}. 
The dataset contained 2,700 and 250 samples for the training and dev sets, respectively. 
Due to the absence of 70 samples in the dataset, the unavailable data were removed during the training phase.

In addition, to train the TTA and TTM data, we utilized the PAM dataset \cite{Deshmukh2023pam}, which contains both natural and generated sounds. 
Moreover, the data were labeled based on the AES listening tests \cite{Tjandra2024audiobox} and are publicly available in the repository\footnote{\url{https://github.com/facebookresearch/audiobox-aesthetics/tree/main/evaluation_data}}. 
For training, we separated the PAM data into training and development sets, each with 900 and 100 samples, respectively, and added them to the AMC25 dataset. 
The main and out-of-domain tracks of the VMC 2022 dataset (VMC22) \cite{huang2022vmc22}, which includes natural, TTS, and voice conversion speech, were also utilized as unlabeled data for training the TTS data. 
From the VMC22 dataset, 8,459 samples were used, with pseudo-labels generated based on the strategy described in the next section. 

Finally, the evaluation dataset for AMC25 was provided with 3,000 synthetic audio samples generated by the TTS, TTA, and TTM systems. 

\subsection{Experimental Conditions}

For the KAN-based model, we set the batch size to 40, number of epochs to 10, and learning rate to $1 \times 10^{-4}$ with a schedule-free version of AdamW as the optimizer \cite{Defazio2024schedulefree}. 
The model was trained for approximately 6--8 GPU hours on an NVIDIA A100 GPU with 80 GB memory. 

For the VERSA-based predictor, we used 28 outputs from existing metrics\footnote{As VERSA is an actively maintained toolkit, we used available scalar-valued and reference-free metrics in the codebase at commit \texttt{c2a7663}.} as input for the regression model. 
The outputs of the regressor correspond to PQ, PC, CE, and CU. We trained each model using 10-fold cross-validation. 
We used Optuna \cite{optuna} to search for the hyperparameters to minimize the MSE of each target.

\subsection{Training Strategies}

For the KAN-based model, we employed a joint strategy of supervised and semi-supervised training using an iterative pseudo-labeling (IPL) approach inspired by \cite{Xu2020ipl,Hwang2022nst}. 
Initially, pseudo-labels were generated using the baseline model as the ``teacher.''
These pseudo-labels were combined with the AMC25 and PAM datasets, which contained the labeled data. 
The ``student'' model is then fine-tuned using this dataset. 
The student replaces the teacher model if the developmental loss is low. 
The process halts when no improvement occurs in the development loss. 
In our experiment, we set the maximum number of updates to five.

The VERSA-based model was trained using AMC25 and PAM datasets.

\begin{table*}[htb]
    \centering
    \caption{Prediction results of the baseline and proposed models fine-tuned with  AMC25 dev-set.}
    \label{tab:results_ft}
    \scalebox{0.785}{
    \begin{tabular}{ccccccc}\toprule
        & &\multicolumn{5}{c}{Utterance-level MSE ($\downarrow$) / SRCC ($\uparrow$)}\\
        \cmidrule{3-7}
        Models & Training Data &PQ & PC & CE & CU    & Overall\\ \midrule
         Baseline&  AMC25&0.590 / 0.806&  0.673 / 0.852&  0.762 / 0.838&  0.719 / 0.818& 0.686 / 0.883\\
         Proposed (KAN)&  AMC25&0.732 / 0.837&  \textbf{0.480} / \textbf{0.895}&  0.859 / 0.853&  0.794 / 0.833& 0.716 / 0.897\\ \midrule
         Baseline&  AMC25 + PAM&\textbf{0.502} / 0.836&  0.589 / 0.883&  \textbf{0.705} / 0.859&  0.669 / 0.841& \textbf{0.617} / 0.899\\
         Proposed (KAN)&  AMC25 + PAM&0.830 / \textbf{0.859}& 0.729 / 0.876&  0.774 / \textbf{0.862}&  0.801 / \textbf{0.850}& 0.783 / \textbf{0.903}\\ \midrule
         Proposed (VERSA)&  AMC25 + PAM&0.535 / 0.833&  0.704 / 0.848&  0.707 / 0.861&  \textbf{0.646} / 0.843& 0.648 / 0.894\\ \bottomrule
    \end{tabular}
    }
\end{table*}

\begin{table*}[htb]
    \centering
    \caption{Prediction results of the proposed models based on the different training strategies with AMC25 dev-set and random seeds \#1--\#4. The FT and IPL denote a fine-tuned and IPL training.}
    \label{tab:results_ipl}
    \scalebox{0.785}{
    \begin{tabular}{ccccccc}\toprule
     & & &  \multicolumn{4}{c}{Overall Utterance-level MSE ($\downarrow$) / SRCC ($\uparrow$)}\\
     \cmidrule{4-7}
      Models & Training Data & Strategies & Seed \#1 & \#2 & \#3 & \#4  \\ \midrule
 Proposed (KAN)& AMC25 + PAM + VMC22 (unlabeled)& FT& 0.574 / 0.908& 0.560 / 0.910&  0.602 / 0.903& 0.595 / 0.904\\
 Proposed (KAN)& AMC25 + PAM + VMC22 (unlabeled)& IPL & \textbf{0.542} / \textbf{0.914}& \textbf{0.557} / \textbf{0.915}& \textbf{0.544} / \textbf{0.913}& \textbf{0.554} / \textbf{0.915}\\ \bottomrule
    \end{tabular}
    }
\end{table*}

{
\tabcolsep = 2.5pt
\begin{table*}[htb]
    \centering
    \caption{Prediction results of the baseline (\texttt{B02}) and proposed system (\texttt{T12}) with AMC25 eval-set, and ablation studies.}
    \label{tab:results_eval}    
    \scalebox{0.785}{
    \begin{tabular}{cccccccccc}\toprule
        & \multicolumn{4}{c}{Utterance-level MSE ($\downarrow$) / SRCC ($\uparrow$)} &  &\multicolumn{4}{c}{System-level MSE ($\downarrow$) / SRCC ($\uparrow$)}\\
        \cmidrule{2-5}
        \cmidrule{7-10}
        Models & PQ & PC & CE & CU   &  &PQ & PC & CE &CU   \\ \midrule
         \texttt{B02} (Baseline)&  1.184 / 0.780&  0.562 / 0.902&  \textbf{1.893} / 0.811&  2.255 / 0.774&  &0.632 / 0.866& 0.226 / 0.934& \textbf{1.142} / 0.841&1.478 / 0.810\\
         \texttt{T12} (KAN \#1--\#4 \& VERSA)&  1.184 / \textbf{0.832}&  0.719 / \textbf{0.911}&  2.472 / \textbf{0.855}&  \textbf{1.853} / \textbf{0.852}&  &0.682 / \textbf{0.916}& 0.401 / 0.938& 1.853 / \textbf{0.946}&1.237 / 0.924\\ \midrule
         Proposed (KAN \#1)&  1.303 / 0.814&  0.751 / 0.906&  2.703 / 0.835&   2.051 / 0.842&  &0.774 / 0.911& 0.407 / 0.936& 2.037 / 0.939&1.331 / 0.935\\
         Proposed (KAN \#1--\#4)&  1.312 / 0.818&  0.746 / 0.909& 2.681 / 0.839& 2.066 / 0.842&  & 0.783 / 0.915& 0.418 / 0.938& 2.031 / 0.941& 1.351 / \textbf{0.939}\\
        Proposed (VERSA)&  \textbf{1.041} / 0.796&  \textbf{0.475} / 0.908&  2.330 / 0.835&   1.923 / 0.811&  &\textbf{0.463} / 0.893& \textbf{0.164} / \textbf{0.946}& 1.596 / 0.911&\textbf{1.099} / 0.908\\ \bottomrule
    \end{tabular}
    }
\end{table*}
}

\section{Results}
Table\,\ref{tab:results_ft} summarizes the prediction results of the baseline and the proposed models by fine-tuning (FT) with the AMC25 and PAM datasets. 
The proposed KAN-based model yielded higher SRCCs for all AES dimensions than the baseline model. 
By adding the PAM dataset, the prediction results improved for both models in terms of the PQ, CE, and CU. 
This indicates that the PAM dataset with AES labels enhances the sound and music information compared with the AMC25 dataset alone.  

The VERSA-based model also demonstrated competitive MSE and SRCC compared with either method for both CE and CU, whereas PQ and PC performed worse. 
We observed that the VERSA-based model predicted outliers in some TTM data, causing a deterioration in evaluation metrics.

Table\,\ref{tab:results_ipl} compares the prediction performance of the KAN-based model trained using the FT or IPL training strategies with that of the unlabeled VMC22 dataset. 
The addition of VMC22 to FT training significantly reduced the MSE; however, the improvement in SRCC was slight. 
Using the IPL training, each of the best KAN-based models was updated approximately three times, and it predicted AESs with better MSEs and SRCCs more stably. 
Therefore, we selected the model trained by IPL training as part of our submitted system. 

Table\,\ref{tab:results_eval} lists the prediction results for the AMC25 eval-set. 
Our submitted \texttt{T12} system, an ensemble model utilizing four KAN-based predictors and one VERSA-based predictor, outperformed the \texttt{B02} (baseline) system in all AES dimensions at the SRCCs. 
For comparison, we evaluated each proposed model: the KAN-based predictor \#1, an ensemble model using four KAN-based predictors, and the VERSA-based predictor. 
The KAN-based ensemble model is comparable and better in terms of the system-level SRCC at the CU, whereas the MSEs are worse than those of the submitted \texttt{T12} system. 
The VERSA-based predictor exhibited better MSEs for all the AES dimensions, whereas the average SRCC was worse than that of the \texttt{T12} system. 
This suggests that using an ensemble model of the KAN- and VERSA-based predictors compensates for the weaknesses of each predictor, particularly for CE and CU. 

Finally, our system achieved the best correlations among the submitted systems for AMC25 Track 2 on PQ, CE, and CU at the utterance level; CE and CU at the system level; and the overall average\footnote{The raw scores and rankings are available on the AMC25 website.}. 
Although the MSEs remained higher than those of the other systems due to the inverse transformation of the target variable and outliers in the predicted scores, this highlights areas for future improvement.

\section{Conclusion}

We proposed an AES prediction system, named AESCA, for the AMC25 Track 2. 
AESCA utilizes a KAN-based audiobox aesthetics predictor and a VERSA-based regression predictor. 
To enhance the generalization performance, the KAN-based predictor was trained using the labeled AMC25 and PAM datasets with the unlabeled VMC22 dataset, based on an IPL strategy. 
Consequently, using the AMC25 evaluation set, our submitted \texttt{T12} system model yielded the best correlations on three axes at the utterance level, two axes at the system level, and the overall average, among the submitted systems. 

\clearpage

\bibliographystyle{IEEEtran}
\bibliography{refs}

\end{document}